\documentclass[twocolumn]{aastex701}

\usepackage{appendix}
\usepackage{hyperref}


\def\gtrsim{\mathrel{\hbox{\rlap{\hbox{\lower4pt\hbox{$\sim$}}}\hbox{$>$}}}}
\def\lesssim{\mathrel{\hbox{\rlap{\hbox{\lower4pt\hbox{$\sim$}}}\hbox{$<$}}}}
\def\farcs{\hbox{$.\!\!^{\prime\prime}$}}

\begin{document}

\title{Pulsed Infrared Emission from Magnetar 4U 0142+61 Detected by JWST}

\author{Jeremy Hare}\email{jeremy.hare@nasa.gov}
\affiliation{NASA Goddard Space Flight Center, Greenbelt, MD 20771, USA}
\affiliation{Center for Research and Exploration in Space Science and Technology, NASA/GSFC, Greenbelt, Maryland 20771, USA}
\affiliation{The Catholic University of America, 620 Michigan Ave., N.E. Washington, DC 20064, USA}
\author{George G. Pavlov}\email{}
\affiliation{Department of Astronomy \& Astrophysics, Pennsylvania State University, 525 Davey Lab, University Park, PA 16802, USA}
\author{Bettina Posselt}\email{}
\affiliation{Department of Astrophysics, University of Oxford, Denys Wilkinson Building, Keble Road, Oxford OX1 3RH, UK}
\affiliation{Department of Astronomy \& Astrophysics, Pennsylvania State University, 525 Davey Lab, University Park, PA 16802, USA}
\author{George Younes}\email{}
\affiliation{NASA Goddard Space Flight Center, Greenbelt, MD 20771, USA}
\affiliation{Center for Space Sciences and Technology, University of Maryland, Baltimore County, Baltimore, MD 21250, USA}
\author{Oleg Kargaltsev}\email{}
\affiliation{Department of Physics, The George Washington University, 725 21st St. NW, Washington, DC 20052}
\author{Alex van Kooten}\email{}
\affiliation{Department of Physics, The George Washington University, 725 21st St. NW, Washington, DC 20052}

\begin{abstract}
We report on a JWST observation of 
the magnetar 4U 0142+61 
on 2024 August 18  
with the Near-Infrared Camera (NIRCam). NIRCam 
observed the magnetar for 33~min in timing mode, providing a time resolution of 2.5~s.
In the F410M filter (pivot wavelength 4.08 $\mu$m), 
we measured the flux density $f_\nu = 22.9\pm0.6$ $\mu$Jy and detected pulsations 
at a frequency 
of $115.059\pm0.035$~mHz, in agreement with the 
magnetar's spin period
at the epoch of the JWST observation. 
The observed
pulse profile has one peak per period (although this may be due to the poor time resolution), with a  lower limit on the pulsed fraction of about 10\%.
 We compare the IR pulse profile to the NICER and NuSTAR X-ray pulse profiles and find that the IR peak overlaps with the hard X-ray peak,
suggesting a magnetospheric origin for the pulsed IR emission.
\end{abstract}

\section{Introduction}
\label{sec:intro}

4U\,0142+61 (4U\,0142 hereafter) is one of the brightest quiescent magnetars  in X-rays \citep{2014ApJS..212....6O}. It has a spin period of about 8.7~s, a magnetic field strength of $1.3\times10^{14}$~G, a characteristic age of 68~kyr, and lies at a distance of 3.6~kpc \citep{2014ApJ...784...37D}. The X-ray luminosity, $L_X\approx10^{35}$ erg s$^{-1}$, of the source in the 2-10 keV band exceeds its spin-down power $\dot{E}=1.2\times10^{32}$ erg s$^{-1}$ by several orders of magnitude. Additionally, the source has exhibited a typical magnetar-like outburst observed by RXTE in 2007 \citep{2008AIPC..983..234G} and by NICER in 2017 \citep{2026ApJS..284...25C}.  In quiescence, it emits pulsed X-ray emission with an RMS pulsed fraction of about $6\%$ at 1 keV increasing to about $20\%$ at 60 keV (see, e.g., Figure 3 in \citealt{2015ApJ...808...32T}). 

4U\,0142 is also one of the brightest magnetars in near-infrared (NIR) and optical (see Table 4 in \citealt{2014ApJS..212....6O}) having a K band magnitude around 20 \citep{2004A&A...416.1037H,2006ApJ...652..576D}. It is one of only three magnetars where pulsations have been detected at optical wavelengths along with  SGR 0501+4516 and 1E1048.1-5937 \citep{2009MNRAS.394L.112D,2011MNRAS.416L..16D}. 
Optical pulsations were originally discovered in 4U\,0142 
in a very broad 
 band by \citet{2002Natur.417..527K}. These authors found a broad pulsed profile with a hint of two peaks per period, resembling the soft X-ray profile but with a factor of 5--10 higher pulsed fraction. However, the optical and X-ray data were not taken simultaneously, and they did not have a phase-connected timing solution, so the phase shift between the optical and X-ray peaks was unconstrained. \cite{2005MNRAS.363..609D} also detected 
pulsations from 4U\,0142 in the $i'$ band,
with an RMS pulsed fraction of $(29\pm8)\%$ and 
a double hump structure (similar to X-rays) in the pulse profile. They 
used a phase-connected timing solution, valid during their optical observations, to determine that the optical peak lags the X-ray peak by $0.04\pm0.02$ cycles.

\cite{2024ApJ...972..176H} reported on observations of 4U\,0142 with JWST using the Mid-Infrared Instrument (MIRI) low-resolution spectrometer (LRS) and the NIRCam 
photometer (JWST observing program \#2635; \citealt{2021jwst.prop.2635P}).
The primary goal was to confirm or refute the existence of a potential fallback disk surrounding the source observed by Spitzer and reported by \cite{2006Natur.440..772W}. The MIRI data showed a featureless absorbed power-law spectrum,
$f_{\nu}\propto\lambda^{\alpha}\propto\nu^{-\alpha}$,
with a spectral index 
$\alpha=0.96\pm0.02$, 
which was further supported by the NIRCam photometry in the F140M and F250M filters. These observations did not support the fallback disk scenario put forth 
by \cite{2006Natur.440..772W}, 
suggesting a non-thermal origin for the 
IR emission instead. 
However, a class of disk models more similar to protostellar disk models \citep{1987ApJ...312..788A}, with a 
temperature dependence different from that suggested by \cite{2006Natur.440..772W} could not be entirely excluded (see Section 3.4 in \citealt{2024ApJ...972..176H} for a detailed discussion). 

\cite{2024ApJ...972..176H} also found that the JWST MIRI fluxes were in good agreement with those previously measured by the Spitzer Infrared Array Camera (IRAC) at 8\,$\mu$m, but there was some discrepancy that was larger than the expected calibration uncertainties between the NIRCam photometry ($<3$\,$\mu$m) and previous observations at NIR wavelengths. For example, HST measured a flux from the source that was about 20\% lower than expected when compared to the best-fit power-law model extrapolated from the JWST data, which far exceeds the calibration uncertainty between the two observatories \citep{2022AJ....163..267G,2024ApJ...972..176H}. Previous studies have suggested that the source is variable at these wavelengths (see, e.g., \citealt{2004A&A...416.1037H,2006ApJ...652..576D}), which may be the cause of the discrepancy, or it may be that the IR and NIR spectral components may come from different emission regions. The picture is further complicated by the fact 
that
the source does not seem to be variable at optical wavelengths (see, e.g., \citealt{2005MNRAS.363..609D,2016MNRAS.458L.114M}) even though the absorbed power-law model fit to the JWST IR and NIR spectral energy distribution (SED) matches the optical data reasonably well when extrapolated to those wavelengths \citep{2024ApJ...972..176H}.

Here we report on new JWST NIRCam observations of 4U\,0142 taken in timing mode. The goal was to search for IR pulsations and compare them to those observed at 
X-ray wavelengths to help further discern the physical processes responsible for the emission. For example, a lag between the IR and X-ray pulse arrival may suggest that the IR emission is produced by a reprocessing of the X-ray emission beyond the magnetosphere of the magnetar (perhaps in a disk). The phase-connected timing solution of \cite{2026ApJ...999...85P}, based on NICER  monitoring of the source 
that covered the time of the JWST observation, is used to compare the pulse profiles at different energies. In Section \ref{sec:obs_and_dat} we discuss the JWST observations and data reduction. In Section \ref{sec:data_analysis} we discuss the time-integrated photometry of 4U\,0142, the phase folding of the IR and X-ray data, and the analysis of the JWST timing data. The discussion of these results is contained in Section \ref{sec:discussion}. Lastly, we summarize our findings and conclusions in Section \ref{sec:summary}.
\\

\section{Observations and Data Reduction}
\label{sec:obs_and_dat}
\subsection{JWST observations}
\label{sec:jwst_obs}

JWST's Near Infrared Camera (NIRCam; \citealt{2023PASP..135b8001R}) observed 4U\,0142 on 2024 August 18. 
A 19.268~s acquisition exposure was
taken in the F335M filter using the SUB32TATS subarray  ($2''\times 2''$ field of view) to ensure the source was placed appropriately on the detector for the time-series observation (TSO). For the 
TSO we used the F410M and F070W filters, with pivot wavelengths of 4.083~$\mu$m and 0.705~$\mu$m and bandwidths of 0.436~$\mu$m and 0.128~$\mu$m, for the long-wavelength and short-wavelength channels, respectively.
The 
TSO was carried out with the SUB64P subarray ($4''\times 4''$ field of view for F410M, and $2''\times 2''$ for F070W filters, with pixels scales of 0.063\arcsec pix$^{-1}$ and 0.031\arcsec pix$^{-1}$ respectively).

The start and end times of the 
TSO were MJD 60540.6272905 
and MJD 60540.6505413, 
which corresponds to a total duration (time span) $T_{\rm span}=2008.88$~s. Here and below the times are Barycenter Dynamical Times, TDB. 
We used the SHALLOW4\footnote{See \url{https://jwst-docs.stsci.edu/jwst-near-infrared-camera/nircam-instrumentation/nircam-detector-overview/nircam-detector-readout-patterns\#gsc.tab=0}.} readout pattern with 10 groups per ``integration ramp'', with each group including 
4 frames (plus 1 skipped frame). In total the observation included $N=800$ integration ramps,
having durations of 
$\Delta t_{\rm ramp}=T_{\rm span}/N=2.5111$ s , with the effective 
exposure time of $\Delta t_{\rm exp} = 2.45784$ s in each ramp (smaller than $\Delta t_{\rm ramp}$ due to the one skipped frame per group). 
This corresponds to the total exposure time $T_{\rm exp}= N\,\Delta t_{\rm exp} = 1966.272$ s.

\subsection{X-ray Observations}
\label{sec:xrayobs}

4U\,0142 has been regularly monitored with 
the NICER observatory \citep{2016SPIE.9905E..1HG} 
since 
2019 March 
with a bi-weekly cadence. Recently, \citet{2026ApJ...999...85P} presented a phase-coherent timing solution 
with a validity range spanning MJD 58504 to 60730 or 2019 January 21 to 2025 February 24. We use this timing solution to phase-fold the NICER and NuSTAR data for comparison with the JWST data. We reduce and clean the NICER data utilizing the \texttt{nicerl2}
tool part of NICERDAS version v15. We cleaned the data by limiting the undershoot rate to be less than 100 and the overshoot rate to be less than 10 and extracted data only from observations that occurred during orbit night. We barycenter-correct the cleaned event files
using the \texttt{barycorr} tool, which is included in HEASoft version
6.36, using the source position from 
\citet{2026ApJ...999...85P} and the JPL planetary ephemeris DE440. Finally, we merged all cleaned event files into one for ease of analysis.

A 41 ks observation using the NuSTAR \citep{2013ApJ...770..103H} 
was taken on 2022 September 20--21 (MJD $\sim59843$), simultaneously with the first JWST MIRI+NIRCam observation. Thanks to the phase-coherent timing solution from \citet{2026ApJ...999...85P}, this dataset can still be used to compare the hard X-ray pulse profiles with those in the IR from JWST. The NuSTAR data were reduced in the same way as described in \cite{2024ApJ...972..176H}, except that we barycentered the event arrival times using the source position from 
\citet{2026ApJ...999...85P} and the JPL planetary ephemeris DE440 before analysis was performed.

\section{Data analysis}
\label{sec:data_analysis}

\subsection{Time-integrated image and photometry}
\label{sec:photometry}
\begin{figure}
\centering
\includegraphics[width=1.0\linewidth]{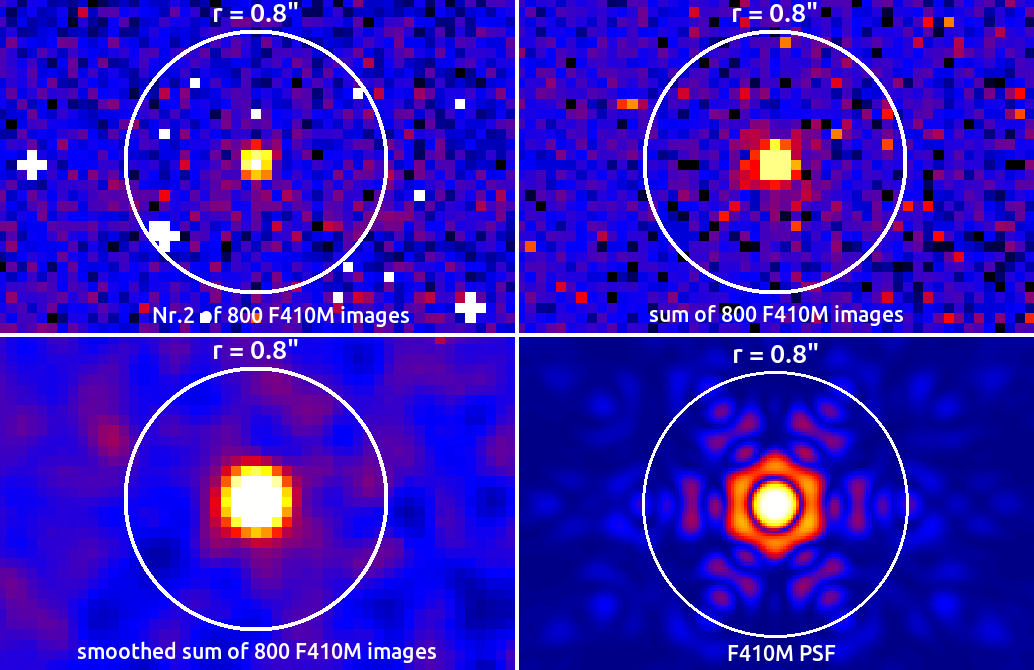}
\caption{The NIRCam F410M TSO data of the target region. Upper left: The second of the 800 integration ramp images.
The white pixels are bad pixels. Upper right: Sum of all 800 ramp images. 
NaN values for bad pixels
are set to 0. 
No smoothing was applied.
Lower left: Same image as upper right, but smoothed. Lower right: WebbPSF for F410M as downloaded from the JWST webpage. No smoothing was applied. Note that 
the four images have different scales, limits, and adapted color scales to highlight the target/PSF region.}
\label{fig:PSF}
\end{figure}

For NIRCam F410M, the image obtained in one of the 800 integration ramps and
the sum of the 800 integration ramp
images are shown in Figure~\ref{fig:PSF}. 
The white pixels 
in the upper left panel of Figure~\ref{fig:PSF} are flagged as `bad' in all integrations. The target's brightness distribution (lower left panel)
was visually compared with the expected point spread function in the F410M filter (lower right panel)
to check for indications of telescope jitter. No indication of such jitter was seen.
We use the image derived from the average of the 800 integration ramps and a corresponding bad pixel mask for our photometry.
To measure the (average) flux density of 4U\,0142, we employ different source aperture sizes centered on the target. 
To subtract the background and evaluate its contribution to the photometric error,
we use the approach of 
``empty background apertures'' (e.g.,  \citealt{Guillot2019,2022ApJ...924..128A,2024ApJ...972..176H}). 
Using an aperture of 0\farcs156 (2.5 pixels) radius with encircled energy fraction (EEF) of 0.65 (see 
the JWST documentation on NIRCam Point Spread Functions\footnote{\url{https://jwst-docs.stsci.edu/jwst-near-infrared-camera/nircam-performance/nircam-point-spread-functions\#NIRCamPointSpreadFunctions-Encircledenergy&gsc.tab=0}}),
 we obtain a $f_{\rm F410M}= 22.9 \pm 0.6$\,$\mu$Jy (see 
  Table~\ref{table:psr_photometry} in Appendix \ref{appendix_photometry} for details).
 
For the F070W,
the summed 800 integrations did not yield a target detection.
We use the source position as well as 26 background positions to estimate upper bounds for the source flux \citep{Guillot2019,Kashyap2010}, 
using multiple aperture sizes, see Table~\ref{table:psr_photometry}. 
We obtain $3\sigma$ upper bounds of 
0.9--1.2~$\mu$Jy, depending on aperture radius. In addition to the NIRCam TSO exposures, we used the 19~s acquisition exposure of 2024 August 18 to measure the flux density at 3.4~$\mu$m. Unfortunately, there is a bad pixel located $\approx0\farcs06$ from the source leading to an unreliable flux measurement, so we excluded it from Figure \ref{fig:new_cal}. Details of the F335M photometry can be found in Appendix \ref{appendix_photometry} (also see Table~\ref{table:psr_photometry}).

\subsection{Re-analysis of the previous spectral fit with updated calibration}
\label{sec:reanalysis}

\cite{2024ApJ...972..176H} noticed changes 
of the best-fit model parameters for the MIRI LRS spectrum of 4U\,0142
caused by changes 
in the instrument's calibration. Furthermore, the calibration of 
NIRCam and MIRI have continued to mature as more data is taken and the instruments are better understood. Therefore, to check if the updated calibration has had any major impacts on the best-fit model, we re-extracted the NIRCam photometry and MIRI LRS spectrum from the 
observations reported in \cite{2024ApJ...972..176H}. We followed the 
procedures outlined in 
that work, using the updated 
calibration version 1.20.2 for MIRI LRS and NIRCam \citep{2025zndo..17515973B}.
We used the pipeline extracted LRS spectrum for MIRI and followed a similar approach for the NIRCam photometry as outlined in the previous section for 
F410M filter. We obtained $f_{\rm F140M}= 5.2 \pm 0.3 \mu$Jy and $f_{\rm F250M}= 14.6 \pm 0.3 \mu$Jy, consistent 
with the previously published values within $1\sigma$.
We did not include the NIRCam F410M 
photometry in this analysis since it was taken at a different epoch and the source may exhibit variability. Note 
that the MIRI LRS and previous NIRCam photometry were taken very close in time.

\begin{figure*}
\centering
\includegraphics[width=0.8\linewidth]{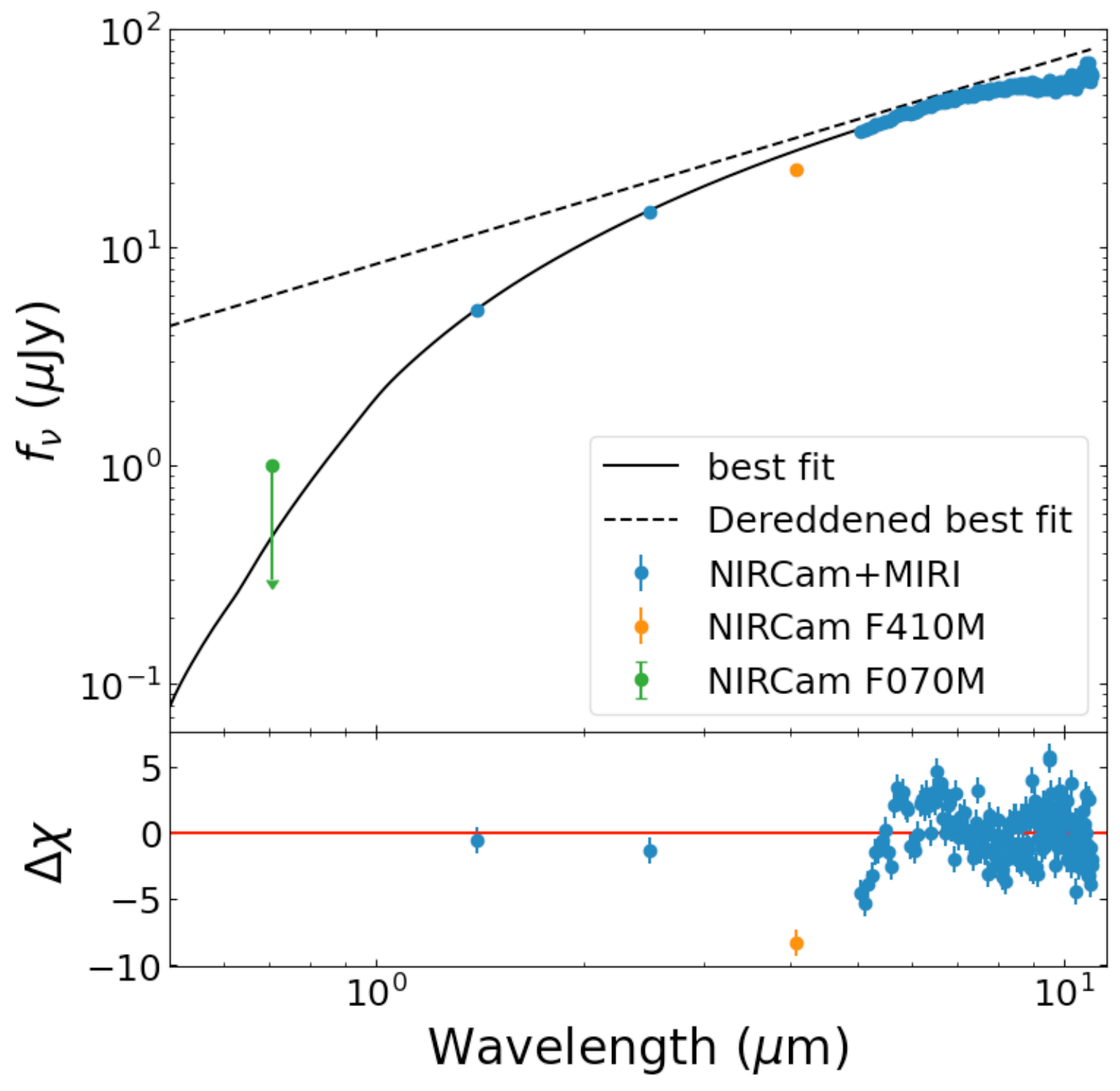}
\caption{Best-fit absorbed power-law model to the MIRI LRS and NIRCam F140M and F250M photometry and corresponding residuals, $\Delta\chi$= (data $-$ model)/error,
obtained with more recent and improved detector calibrations. Note that the F070W and F410M 
fluxes
were not used in this fit as the data were obtained at different epochs. The fit is in good agreement with that from  \citealt{2024ApJ...972..176H}. The dereddened best-fit model is shown by the dashed line. The error bars 
in the bottom panel 
correspond to $\pm1\sigma$. 
}
\label{fig:new_cal}
\end{figure*}

We jointly fit the MIRI LRS spectrum and NIRCam photometry from the F140M and F250M filters with an absorbed power-law, $f_\nu = f_0 (\lambda/\lambda_0)^{\alpha} 10^{-0.4 A_\lambda}$. We find best-fit values of $A_V=3.9\pm0.1$, $f_0=60.1\pm0.3$\,$\mu$Jy at $\lambda_0=$8$\,\mu$m, and $\alpha=0.95\pm0.01$ with a reduced chi-squared $\chi^2_{\nu}=4.0$ (see Figure \ref{fig:new_cal}). These values are in good agreement with 
 $A_V=3.9\pm0.2$, $f_0=59.4\pm0.5$ $\mu$Jy, and $\alpha=0.96\pm0.02$,
found by \citet{2024ApJ...972..176H}, suggesting that 
the updated calibration does not have a large impact on the best-fit model parameters. 
The new reduced chi-squared is about a factor of 2 lower than found by \citet{2024ApJ...972..176H}, and is mainly due to larger, more realistic error bars on the MIRI LRS spectral data points. However, it still suggests that the errors are somewhat underestimated. Therefore, we increase the uncertainties by a factor of two, which gives $\chi^2_{\nu}=0.99$, to calculate the errors in the best-fit parameters reported above in this paragraph. The data, best-fit model, and F410M photometry are shown in Figure \ref{fig:new_cal}. It is clear that the F410M filter photometry is significantly below the interpolation of the best-fit power-law model by about 5 $\mu$Jy, or $\sim20\%$ (see Section \ref{sec:discussion} for further discussion).

\subsection{Timing analysis of the NIRCam data}
\label{sec:timing_analysis}
\begin{figure}
\centering
\includegraphics[width=1.0\linewidth]{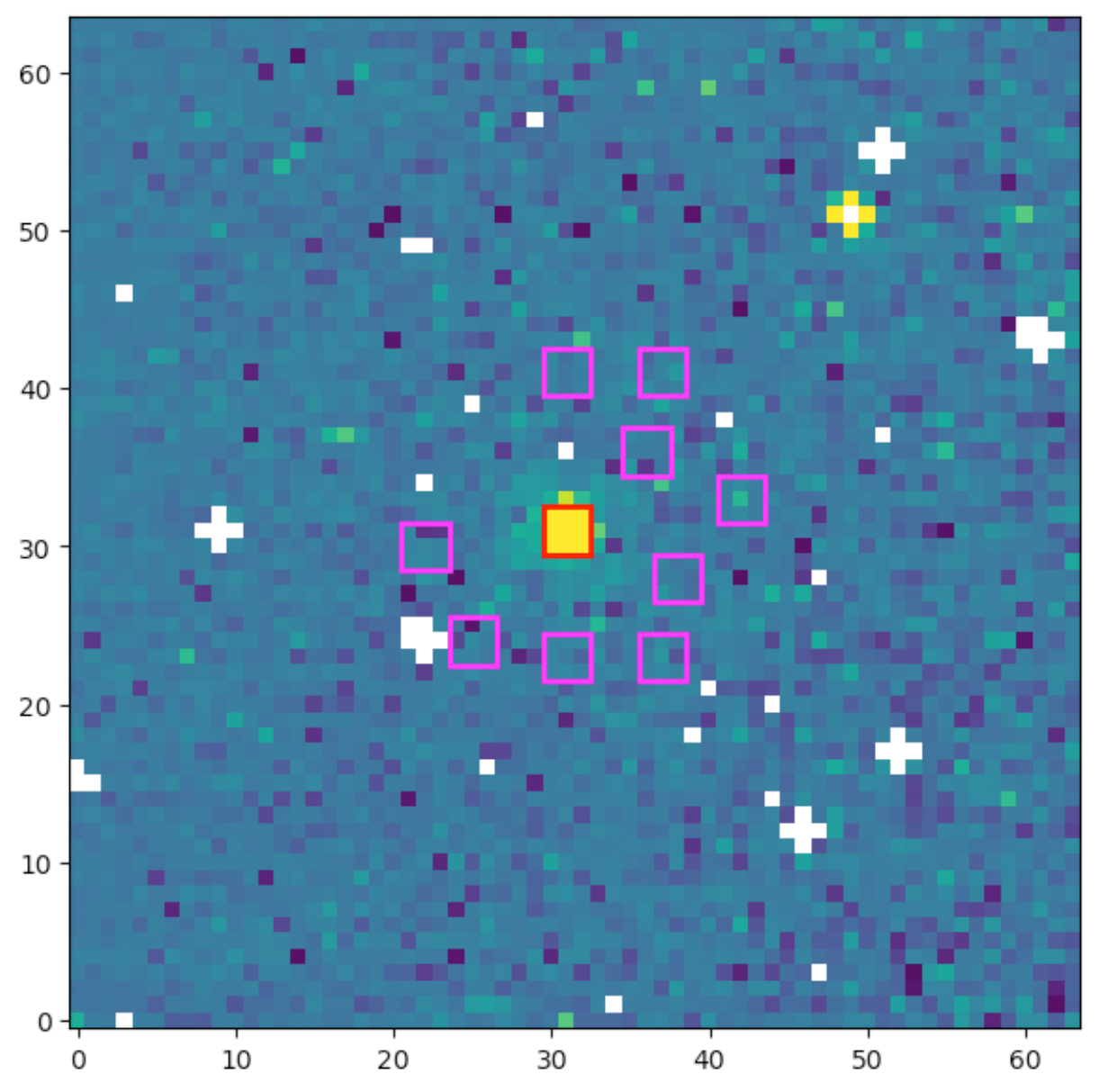}
\caption{F410M filter image from a single 
integration ramp of the timing observation of 4U\,0142. The white pixels correspond to known bad pixels which are masked in the calibrated images.
Red shows the source region, and magenta shows the nine background regions used for the analysis.}
\label{fig:regions}
\end{figure}

We extracted the light curves from a 
$3\times 3$ pixels ($0\farcs189\times 0\farcs189$) source region centered on 4U\,0142. Nine $3\times3$ pixels 
squares offset from the source were chosen for the background evaluation (see Figure \ref{fig:regions}).
The net source light curve was obtained by subtracting the mean of the background light curves
from the light curve of the source region. Each of the light curves is given by flux density values in the $3\times 3$ pixels regions 
in $N=800$ equal time intervals $\Delta t_{\rm ramp} = 2.5111$\,s, with the total time span $T_{\rm span} = N\,\Delta t_{\rm ramp} =2008.88$\,s. The source dominates the background,
0.4$\pm0.2$ $\mu$Jy 
per $3\times 3$ pixels region, with a mean time-averaged background-subtracted source flux of $\bar{f}_\nu \pm \delta \bar{f}_\nu = 12.7\pm0.2$ $\mu$Jy, which is about 55\% of the total (aperture-corrected) flux of the magnetar at 4.1 $\mu$m. We checked that this aperture maximizes the signal in the timing analysis reported in Section \ref{sec:ir_power_spectrum}.

\subsubsection{Power spectrum of the NIRCam data}
\label{sec:ir_power_spectrum}
\begin{figure}
\centering
\includegraphics[width=1.0\linewidth]{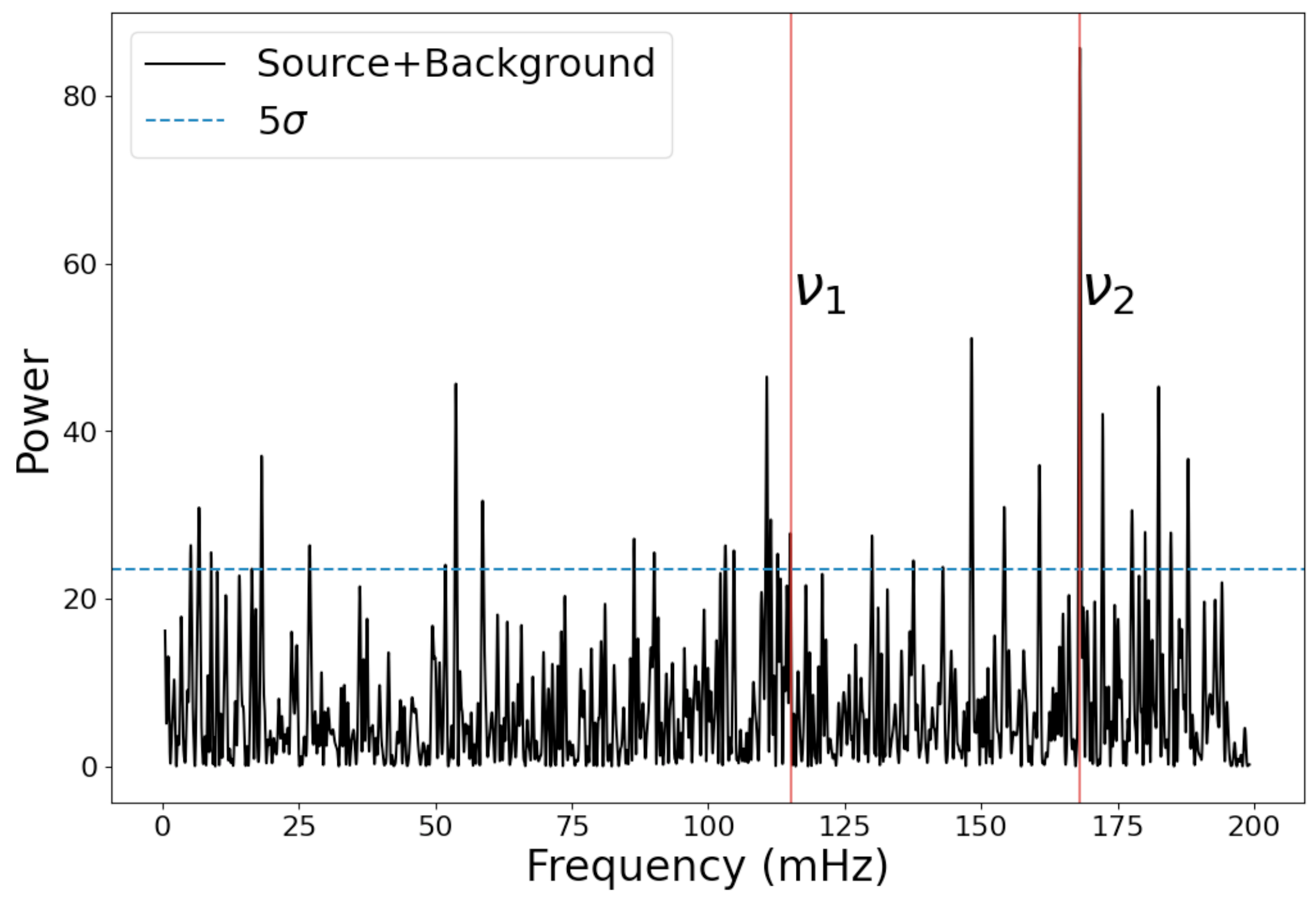}
\includegraphics[width=1.0\linewidth]{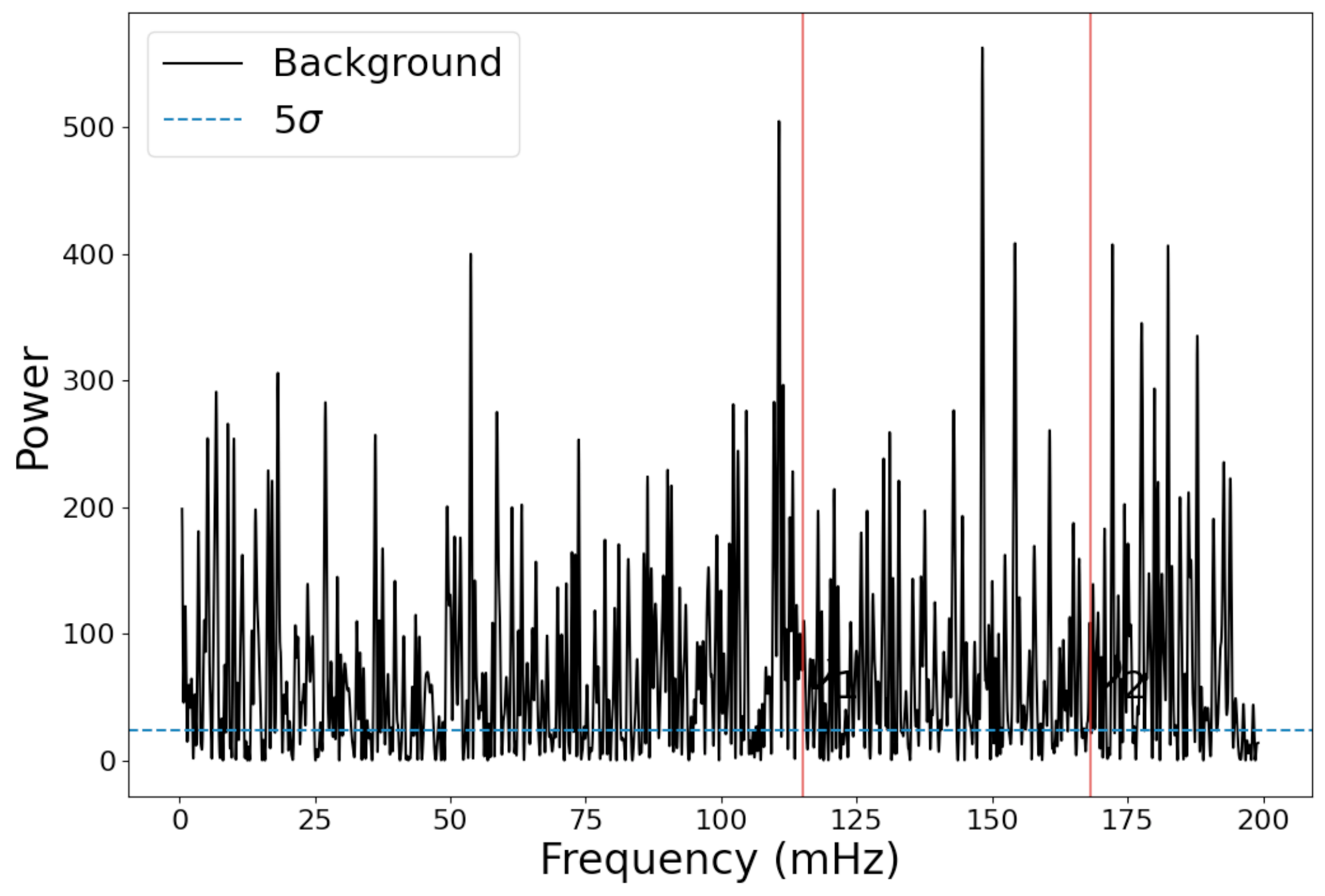}
\includegraphics[width=1.0\linewidth]{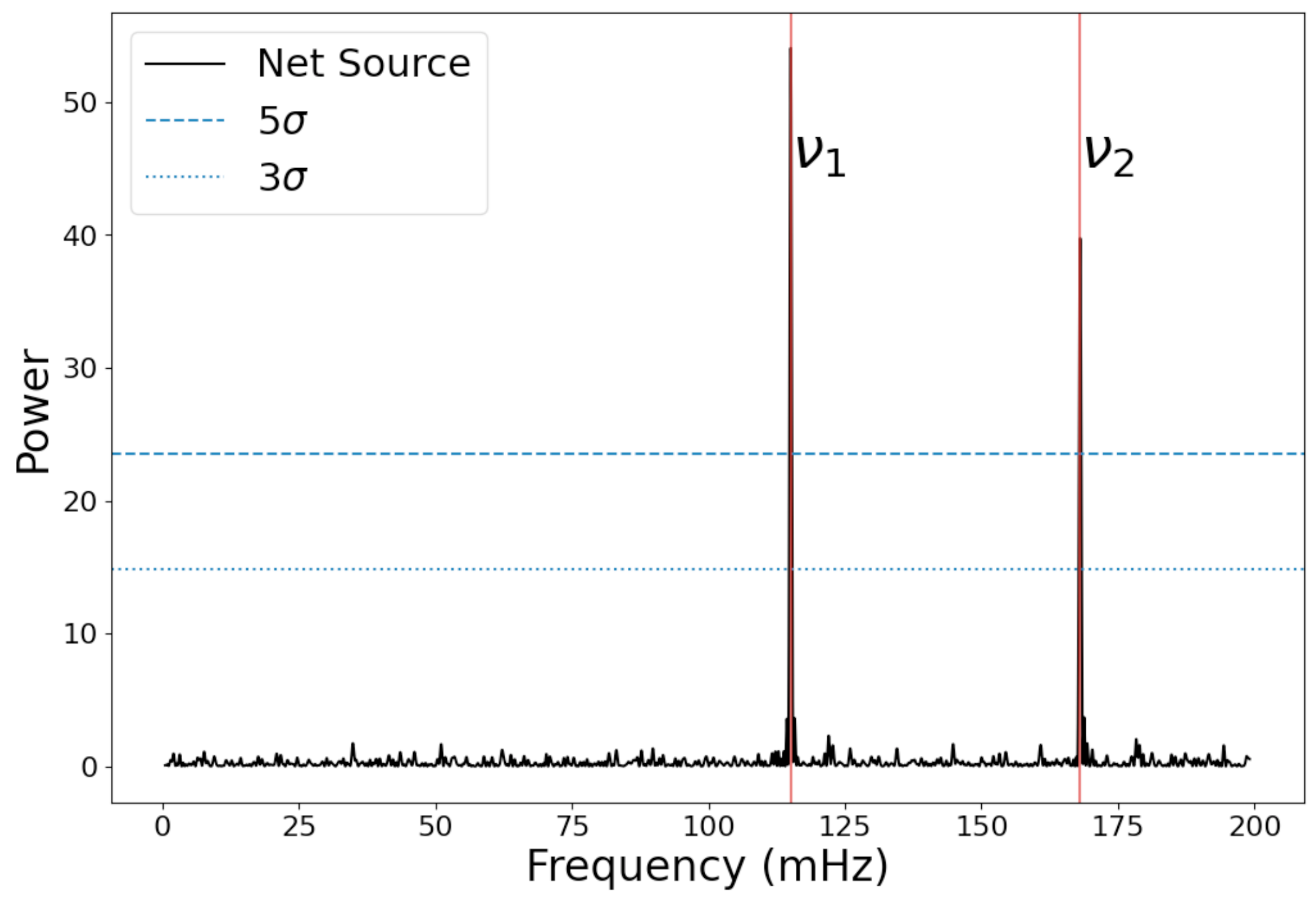}
\caption{Lomb-Scargle periodograms of the source region without
background subtraction (top), mean of the background regions (middle), and source region with background subtracted (bottom). 
The red solid lines show the magnetar spin peak at $\nu_1\simeq 115$ mHz and the aliased peak $\nu_{2}\simeq 168$ mHz.
The blue horizontal dotted and dashed lines in the bottom panel shows the power corresponding to a 3$\sigma$ and 5$\sigma$ trials-corrected significance, respectively. Only the 5$\sigma$ trials-corrected significance is shown in the top and middle panels.
}
\label{fig:power_spectra}
\end{figure}

We calculated the Lomb-Scargle (LS) periodogram to search for the spin period of the magnetar using the astropy implementation of the LS algorithm \citep{2018ApJS..236...16V,2022ApJ...935..167A}. The barycenter-corrected times from 
the mid-point of each light curve time bin, provided in the calibrated JWST files, were used to construct the power spectrum. The length of the observation sets the minimum frequency 
of $\approx0.5$ mHz, while the $t_{\rm samp} = t_{\rm ramp} =2.5111$~s
sampling time gives a Nyquist frequency $\nu_{\rm Nyq} = (2t_{\rm samp})^{-1} = 199.116$~mHz, so we searched for pulsations in this frequency range.
We used a step size of $\delta\nu=(100 T_{\rm span})^{-1}\sim0.004978$ mHz, which includes an oversampling factor of 100, corresponding to a total of roughly 40,000 frequencies searched. We calculated the power spectrum, shown in separate panels of  Figure \ref{fig:power_spectra}, using 
the source light curve without subtracting the background, the mean (across the 9 regions) background light curve, and the background-subtracted source light curve.

\begin{figure}
\centering
\includegraphics[width=1.0\linewidth]{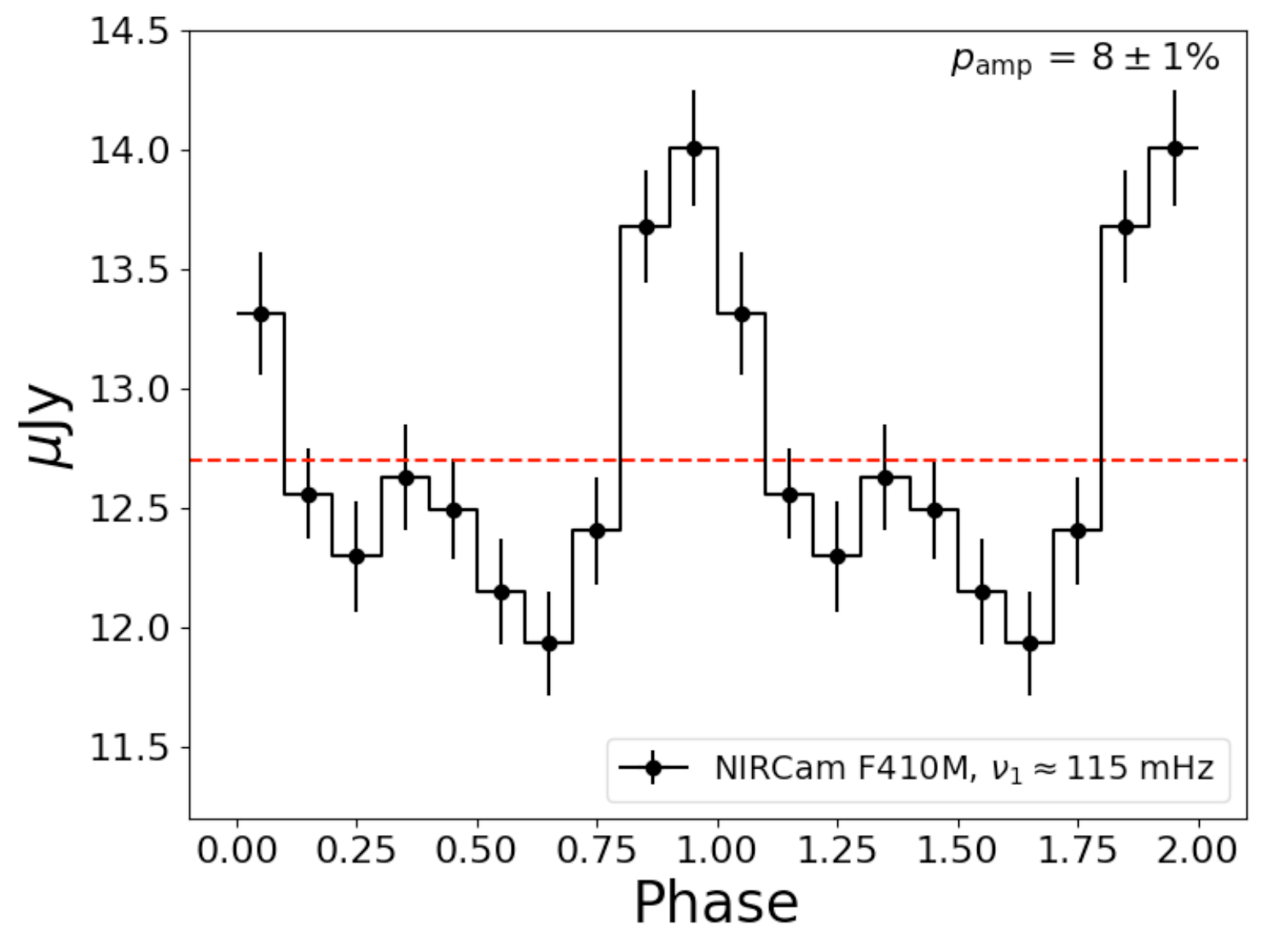}
\caption{Binned phase-folded NIRCam light curve for $\nu =\nu_1$ (i.e., the spin period of the magnetar). 
The reference time corresponding to phase 0 is MJD 60540.63881582. 
The red dashed line shows the time-averaged flux of the source.
}
\label{fig:ir_lc}
\end{figure}

The power spectrum of the background-subtracted light curve shows two 
high peaks 
at frequencies
${\nu_1=115.059\pm0.035}$ mHz and ${\nu_{2}=168.079\pm0.035}$ mHz, 
with unnormalized LS powers of 54.1 and 39.7, respectively (see Figure \ref{fig:power_spectra}).
The highest peak 
lies close to the anticipated 
frequency
of the magnetar pulsations (associated with the NS spin),  $\nu_{\rm spin} =115.07724616(4)$ mHz
at the mid-point of the JWST observation (i.e., MJD 60540.6389159), calculated from the timing model of \cite{2026ApJ...999...85P} using the Code for Rotational-analysis of Isolated Magnetars and Pulsars (CRIMP\footnote{https://github.com/georgeyounes/CRIMP}).
The offset between these frequencies is  
$0.018\pm 0.035$~mHz,
 i.e., $\nu_1$
coincides with 
$\nu_{\rm spin}$ within statistical uncertainties.

Although one might speculate that the second highest peak is caused by another periodic process in the magnetar (e.g., a small object orbiting the neutron star on a $\sim 10^9$ cm radius orbit), a more natural explanation is that it is an alias connected with data sampling and the presence of second harmonic in the signal.
For a sampling frequency $\nu_{\rm samp} = (t_{\rm samp})^{-1} = 2\nu_{\rm Nyq}$ and a pulsed signal from the magnetar that contains several harmonics at frequencies $k\nu_{\rm spin}$ ($k=1$, 2, \ldots), one can expect aliases at frequencies
$\nu_{a,k} = \nu_{\rm samp} - k \nu_{\rm spin}$ 
(see, e.g., Eq.\ (46) in \citealt{2018ApJS..236...16V}).
The frequency of the $k=1$ alias, $\nu_{a,1}=283.155$~mHz, lies above the Nyquist frequency, but 
$\nu_{a,2}=168.077$~mHz coincides with the frequency $\nu_2$
with a negligibly small offset of $0.002\pm 0.035$~mHz. Thus, the presence of this alias peak in the power spectrum implies the presence of a second harmonic in the IR pulsations.

We carried out some additional checks to further substantiate this hypothesis. We first checked that the peak at this frequency (as well as the peak at the spin frequency $\nu_1\simeq 115$ mHz) is 
seen in the source region in 400 ramps ($\approx 1000$ s) sub-samples of the entire data
set. This proves that
both the 115 mHz and 168 mHz peaks are persistent throughout
the entire observation. To check 
 that the 168 mHz pulsations are connected with the magnetar (i.e., an alias of the spin period),  we treated 
one of the 9 background squares as the source region and subtracted the background determined from the remaining 8 background squares. 
The background-subtracted LS periodogram did not show a 
detectable signal at
168 mHz, further supporting that the signal is an alias of the magnetar's spin frequency.
Finally, we checked that the alias $\nu_{a,1}$ is indeed seen in the power spectrum at 283.163 mHz, which also has a negligibly small offset of 0.008 mHz.

\subsubsection{Folded light curves and pulsed fractions of the IR pulsations}
\label{sec:ir_light_curves}

Figure \ref{fig:ir_lc} shows phase-folded light curves at the frequency $\nu_1$ measured by the NIRCam observations.
In the phase-folded pulse profile 
we see one peak per period, with a width 
$\delta\phi \approx 0.3$ ($\delta t\approx 2.6$ s).
Using this pulse profile, we calculate a ``peak-to-trough'' 
pulsed fraction\footnote{See, e.g., \cite{2021ApJ...923..249H} for different definitions of pulsed fractions and their relation to each other.} 
 $p_{\rm amp} = 8\%\pm1\%$ (or an ``RMS pulsed fraction'' 
$p_{\rm RMS} = 6\%\pm1\%$).

It is important to note, however, that the finite time resolution
suppresses the amplitude and power in a $k$-th harmonic of
a signal by factors $\xi_k = {\rm sinc} (\pi k \nu t_{\rm ramp})$ and $\xi_k^2$, respectively (see Appendix \ref{sec:finite_timeres}), which distorts the pulse profile and reduces the pulsed fraction. At $t_{\rm ramp} = 2.51$ s, as in the NIRCam timing observation, the suppression factors for the first two harmonics are $\xi_1= 0.87$ and
 $\xi_2= 0.54$.
Since the suppression of the pulsed fraction depends on the unknown relative strengths and phases of the signal harmonics, we cannot quantitatively correct it for the finite time resolution, but we can 
expect the true pulsed fraction to exceed the above values (e.g., $p_{\rm amp} > 10\%$).

\subsection{Comparison of the IR and X-ray Pulse profiles}
\label{sec:IR_vs_xrays}

To compare the discovered IR pulsations with the X-ray pulsations, we need an X-ray timing solution applicable at the epoch MJD 60540 of the NIRCam timing observation.
A phase-connected timing solution has been provided by \cite{2026ApJ...999...85P}, who analyzed X-ray pulsations detected over 6 years of NICER monitoring of 4U 0142 between MJD 58504.0--60730.1. This timing solution has RMS residuals of $\sim61$ ms, or 0.7$\%$ of the magnetar's spin period, with the maximum TOA residuals being offset by about 0.03 in phase. This solution shows two spin-down glitches, at the epochs MJD 59540 and 59850, with frequency jumps of $\Delta\nu_{g1} = -1.25(9)\times 10^{-8}$ Hz and $\Delta\nu_{g2}=-5.42(9)\times 10^{-8}$ Hz, respectively. Coincidentally, the second glitch occurred about 7 days after our previous JWST and NuSTAR observations.
In Tables 5 and 6, \citet{2026ApJ...999...85P} provide both the 6-year solution with glitches included (Table 5 of that paper) and separate solutions in three time segments: 
before the first glitch (S1),
between the glitches (S2),
and after the second glitch (S3).
These authors noticed slight (but statistically significant) differences of NICER pulse shapes in different segments (see their Figure 5), but no pulse variations were noticed within the segments. Therefore, we use the global timing solution\footnote{Note that there is a sign error in the $\Delta\phi_{g1}$ and $\Delta\phi_{g2}$ in Table 5 of \cite{2026ApJ...999...85P}, confirmed by private communication with \cite{2026ApJ...999...85P}, which we correct for when phase folding our data.} to phase fold the NICER data using the PINT python package \citep{2021ApJ...911...45L,2024ApJ...971..150S}. 

Given that there were differences in the pulse shapes between segments, we split the data into two epochs. Epoch 1 spans the MJD range after the first glitch from MJD 59540 to 59850, while Epoch 2 spans the MJD range after the second glitch from MJD 59850 to 60726. Note that the NuSTAR and JWST time series observations occurred in Epoch 1 and Epoch 2, respectively. The reference time corresponding to phase 0 of the \cite{2026ApJ...999...85P} timing solution, closest to the midpoint of the JWST TSO, 
is MJD 60540.63881582 and was calculated using CRIMP.

In the comparison between the IR and X-ray pulsations, we need to take into account the phase precision of the corresponding pulse profiles. 
The absolute timing accuracy of JWST has been shown to be on the order of 0.1 s \citep{2025AJ....169...21S}, corresponding to $\sim 0.01$ of the magnetar's spin period. However, the poor time resolution of JWST
has a much larger impact on determining the precise phase where the IR pulse profile peaks. 
The precision with which the NIRCam pulsation phase can be measured is estimated as 
$\sigma_\phi \sim w_\phi (S/N)^{-1} \approx 0.064$,
where $w_\phi = \nu_1 \Delta t_{\rm ramp} \approx 0.3$ is the phase resolution ($\Delta t_{\rm ramp}\simeq 2.5$ s is the time resolution), $S/N = p (\bar{f}_\nu/\delta \bar{f}_\nu) \approx 4.7$
is the signal-to-noise ratio for the pulsed flux ($\bar{f}_\nu$ and $\delta\bar{f}_\nu$ are the time-averaged flux density and its uncertainty in the $3\times 3$ pixel region chosen for the analysis; see Section \ref{sec:timing_analysis}), and $p$ is the pulsed fraction. This gives a rough estimate of 0.064 for the determination of the phase of the IR pulse peak. However, given that the pulse shape is not sinusoidal, and 
the poor time resolution can wash out the pulse structure (e.g., the double-peaked pulse profile),
the true uncertainty may be larger.

\begin{figure*}
\centering
\includegraphics[width=1.0\linewidth]{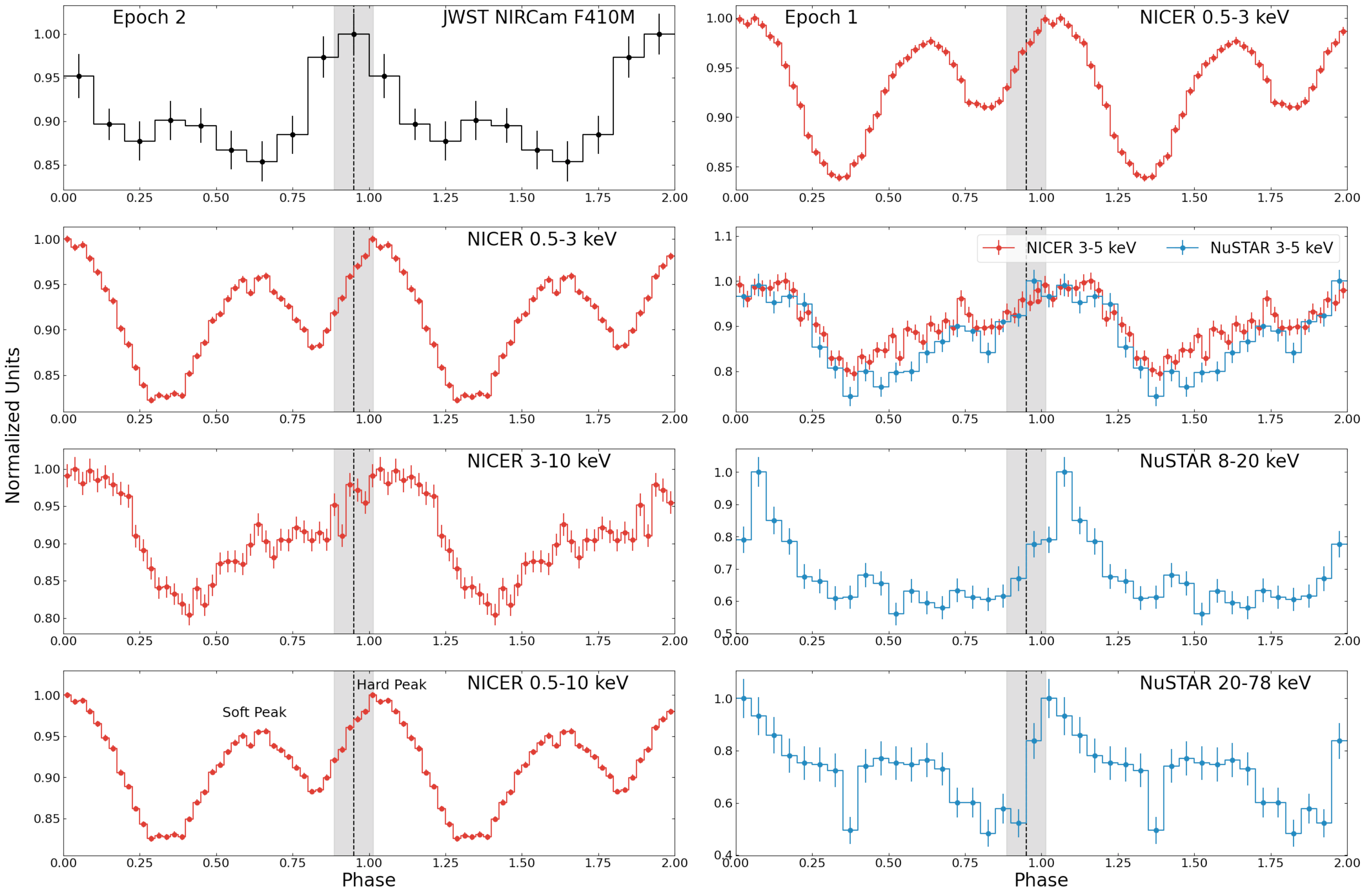}
\caption{Binned phase-folded light curves for JWST NIRCam, NICER, and NuSTAR observations using the timing solution presented in Table 5 of \cite{2026ApJ...999...85P}. The four left panels show the NIRCam pulsations, detected in a $\approx2$ ks observation taken on MJD 60540.64, together with NICER pulsations in 3 energy bands from Epoch 2, spanning MJD 59850 to 60726.
The four right panels show NuSTAR pulsations in 3 energy bands, observed on MJD 59843, together with NICER pulsations in 2 energy bands observed in Epoch 1, spanning MJD 59540 to 59850.
 The dashed black line shows the phase corresponding to the maximum of the IR pulse profile with the gray band  being an estimate of the uncertainty. The reference time corresponding to phase 0 is MJD 60540.63881582. 
}
\label{fig:full_folded_lc}
\end{figure*}

Figure \ref{fig:full_folded_lc} shows the folded NICER and IR pulse profiles from Epoch 2 (left panels) and the NICER and NuSTAR pulse profiles from Epoch 1 (right panels). The X-ray pulse profile shows two strong peaks in the 0.5--10 keV energy band around
phases 0.1 and 0.66.
The first peak,
dominated by the hard power-law tail of the magnetar's spectrum, persists throughout the entire X-ray energy range of 0.5--78 keV, changing shape with energy.
The second peak is most prominent in the 0.5--3 keV energy range and virtually disappears in the 8--20 keV band. 
It is associated with 
thermal emission from a heated region of the neutron star surface, with a contribution from a soft power-law spectrum (see, e.g., \citealt{2015ApJ...808...32T}). Previous studies have found that the RMS pulsed fraction of the X-ray emission grows with energy, from $\sim5\%$ at 1 keV to $\sim20\%$ above 20 keV \citep{2015ApJ...808...32T}, consistent with the strong pulsations of the hard peak at high energies shown here. The IR peak, along with a rough estimate of the uncertainty of the peak phase due to the poor time resolution, is shown as a gray band in all the panels of Figure \ref{fig:full_folded_lc} .

\section{Discussion}
\label{sec:discussion}

The timing analysis of the NIRCam F410M data 
enabled the detection of pulsations with the magnetar's spin frequency, $\nu_{\rm spin}\simeq 115$ mHz, known with a much higher precision from X-ray observations (see Section \ref{sec:IR_vs_xrays}).
In the phase-folded light curve (Figure \ref{fig:ir_lc})
we  
see one broad peak per period,
 with a width $\delta\phi \approx 0.3$ and a pulsed fraction $p_{\rm amp}=8\%\pm 1\%$.
We should note, however, that the shape 
of these pulsations is distorted
by the poor time resolution of our NIRCam observation, $\Delta t_{\rm ramp} \simeq 2.51\,{\rm s} \simeq 0.29 P_{\rm spin}$, so that we cannot exclude the possibility that more than one peak would be seen if observed with better time resolution. Moreover, the poor time resolution suppresses the pulsations, which means that the intrinsic pulsed fraction may be substantially larger than the measured value.

Some hints as to the nature of the discovered IR pulsations may be provided by comparing them to the well-studied X-ray pulsations of 4U\,0142 (see, e.g., \citealt{2015ApJ...808...32T}, and references therein). The IR peak comes slightly before the hard X-ray peak (see Figure~\ref{fig:full_folded_lc}),
with 
phase offsets of about $-0.063$, $-0.125$, and $-0.075$
from the NICER peak in the 0.5--10 keV band and NuSTAR peaks
in the 8--20 and 20--78 keV,
 bands, respectively. These peak phases are consistent within the (1--2)\,$\sigma$ uncertainties of the IR peak phase. To more reliably measure the relative phases of the X-ray and IR pulsations and suggest a detailed model, IR observations with a significantly better time resolution are required. However, if the IR peak does arrive before the hard X-ray peak (as hinted at in Figure \ref{fig:full_folded_lc}), it suggests that the IR and hard X-ray emission, even if connected with each other, 
are produced at different emission sites. This is further supported by the fact that the IR and hard X-ray spectra do not smoothly connect (see \citealt{2024ApJ...972..176H}). 

According to popular magnetar models,
the nonthermal X-ray component of a magnetar's persistent emission is produced by relativistic electrons and positrons moving from the NS surface along a loop of twisted magnetic field lines and losing their energies to the resonant Compton upscattering of thermal soft X-ray photons
(see, e.g., Figure 8 in \citealt{2017ARA&A..55..261K}). The height of this loop is likely $\lesssim 10 R_{\rm NS}$. The mechanism and site(s) of the magnetar's IR-optical emission are currently unclear. In principle, the IR-optical emission could be interpreted as coherent curvature radiation from the inner magnetosphere, at distances $\sim 10 R_{NS}$ from the neutron star surface \citep{2011ASSP...21..329Z}. However, other possibilities are not excluded.
For instance, one can imagine that some originally relativistic electrons and positrons move along open magnetic field lines,
decelerate down to MeV energies and produce IR-optical photons via synchro-cyclotron radiation at much larger distances from the NS. Future near simultaneous IR-optical observations (to minimize the effects of variability) and more detailed modeling of the IR-optical emission from magnetars are needed to better understand the IR-optical emission and its possible relation to the X-ray emission.

As mentioned above, the IR 
pulsed fraction, corrected for the time resolution, should be higher than the measured $p_{\rm amp} = 8\%\pm1\%$.
The true pulsed fraction may be similar to $29\%\pm 8\%$, reported by \cite{2005MNRAS.363..609D} from observations of 4U\,0142 in the $i'$ (0.8\,$\mu$m)  band, exceeding the X-ray 
$p_{\rm rms}\lesssim 20\%$. If such a high IR pulsed fraction is confirmed by future observations, this would be an additional argument against the hypothesis that the IR emission is caused by reprocessing of the magnetar's X-ray emission by some circumstellar material (such as a fallback disk).

Another result of this work
is the possible variability of the IR flux. As shown in Figure \ref{fig:new_cal} and Sections \ref{sec:photometry} and \ref{sec:reanalysis}, the flux density value measured with F410M  
on 2024 August 18 lies $\sim$20$\%$ below the best-fit spectrum measured with MIRI LRS and NIRCam F140M and F250M filters on 2022 September 21. Assuming the same spectral model as reported in Section \ref{sec:reanalysis} and renormalizing it based on this $\sim$20$\%$ lower flux value in the F410M filter gives an unabsorbed luminosity $L_{\rm 1.4-11 \mu m}\simeq6.1\times10^{31}$ erg s$^{-1}$ (versus $7.3\times 10^{31}$ erg s$^{-1}$ in the previous observation) at a distance of 3.6 kpc. 

\begin{figure}
\centering
\includegraphics[width=1.0\linewidth]{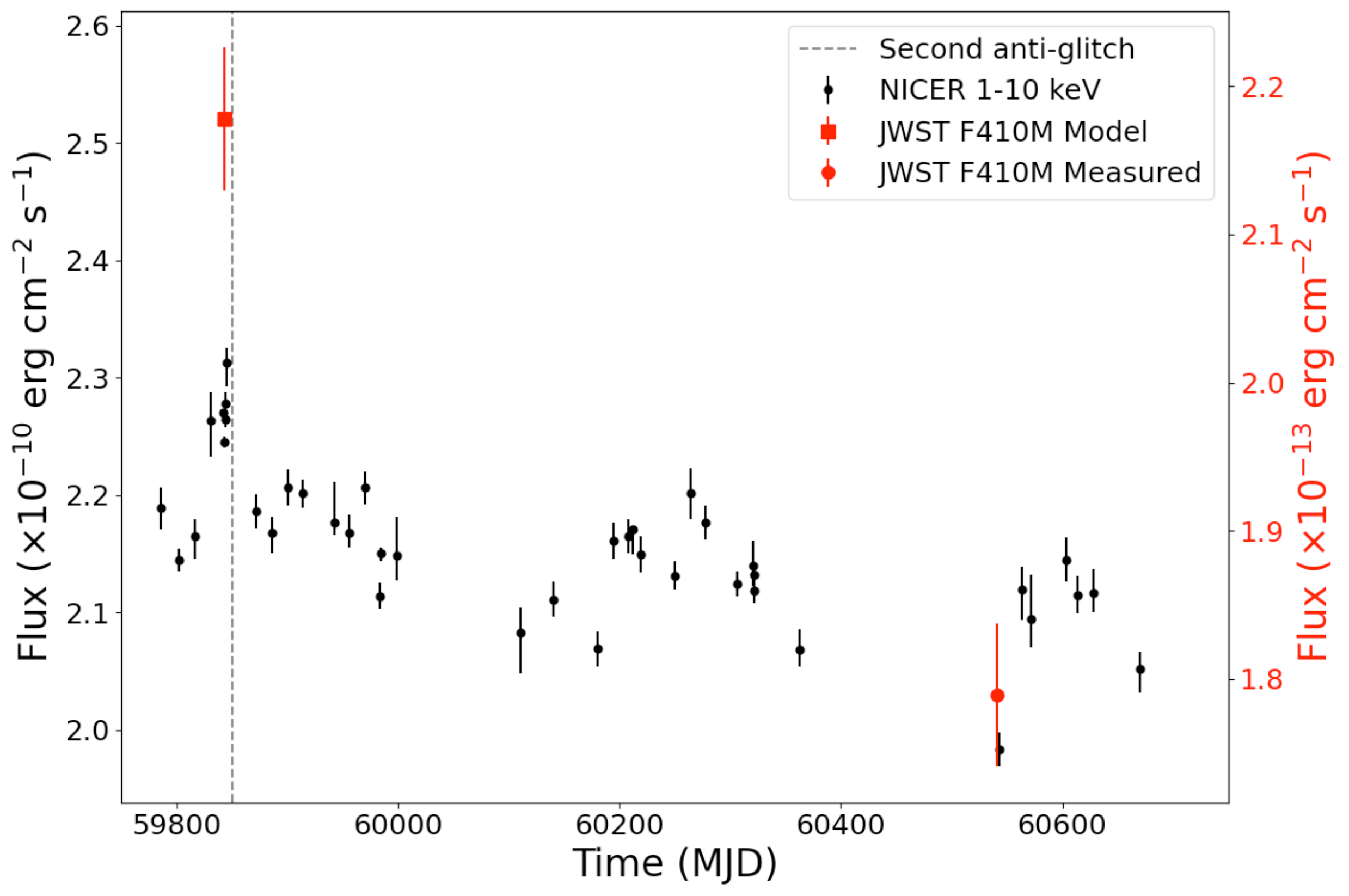}
\caption{ NICER 1-10 keV X-ray (black) and NIRCam F410M (red) flux of 4U 0142+61 versus time. The time of the second glitch reported by \cite{2026ApJ...999...85P} is shown as the dashed gray vertical line. See Section \ref{sec:discussion} for details on how the F410 model flux was derived.
}
\label{fig:X_ray_IR_flux}
\end{figure}

4U\,0142 has shown X-ray bursts around the time of glitches, causing the flux to increase and then decay (see, e.g., \citealt{2017ApJ...834..163A}). To determine if such a burst occurred near the second anti-glitch reported by \cite{2026ApJ...999...85P},  we extracted the unabsorbed X-ray flux of 4U\,0142 from NICER observations\footnote{ Details of how the fluxes were extracted will be presented in a forthcoming paper (Van Kooten in prep.).} spanning the time interval between the JWST observations (see Figure \ref{fig:X_ray_IR_flux}). We also overplot the unabsorbed NIRCam fluxes from the two JWST observations. Note, however, that the F410M filter was not used in the first JWST observation, so we estimated its flux value from the best-fit model reported in Section \ref{sec:reanalysis}. The error bar was set to the same value (0.6 $\mu$Jy) as the measured F410M photometry reported in Section \ref{sec:data_analysis}. A small increase in X-ray flux was observed around the time of the second anti-glitch and we find a correlation between the X-ray and IR fluxes, with an X-ray to IR flux ratio of about 1000, which remains consistent between the two JWST observations. Higher cadence monitoring of the source in the IR is needed to confirm this correlation, but in any case, the change in IR flux may be related to the increase in X-ray flux near the time of the anti-glitch.

There have been several models put forth attempting
to explain the IR-optical emission from magnetars with disks of gaseous or dusty matter rotating around these objects, such as fallback disks formed 
in the aftermath of the supernova explosion. 
Different disk models may be able to account for some of the observed IR-optical properties, such as the pulsations (see e.g., \citealt{2004ApJ...605..840E}) and long-term variability. For instance, \citet{2007ApJ...657..441E} discussed an active irradiated disk for 4U\,0142, which could, in principle, account for variability (e.g., due to changes in the mass-flow rate\footnote{Such changes may also cause torque variations, which would show up in the timing behavior of the source.}, inner disk radius,  irradiation by X-ray outbursts, or disk instabilities). However, so far a comprehensive disk model 
that can account for all of the observed IR/optical properties, including a broad power-law spectrum, long-term variability, and pulsations that are nearly aligned with the hard X-ray pulsations is still lacking. 
Moreover, it seems very plausible that the observed IR-optical properties of 4U\,0142, including an almost flat ($\alpha=0.95$) power-law spectrum from IR to optical (although not simultaneously observed) and pulsed IR emission with an emission peak virtually coinciding with the hard X-ray peak are consistent with a non-thermal (magnetospheric) origin and can be explained without invoking a disk.

\section{Summary}
\label{sec:summary}

The new JWST NIRCam observation of the magnetar 4U\,0142+61 allowed us to discover pulsations of its radiation at wavelengths around 4.1 $\mu$m at the magnetar's spin frequency $\nu_{\rm spin}\simeq 115$ mHz. The IR pulse overlaps with the hard X-ray pulse. This provides further support to the conclusion, based on spectral analysis \citep{2024ApJ...972..176H}, that the magnetar's IR-optical emission is of a nonthermal origin, instead of thermal emission from a fallback disk as suggested by  \citet{2006Natur.440..772W}. 

The magnetar's flux density in the F410M filter, $f_\nu =22.9\pm 0.6$ $\mu$Jy, is 20\% lower than expected from the spectrum measured by \citet{2024ApJ...972..176H} two years earlier, which suggests that the IR emission is variable.

The reported exploratory timing observation was too short and done with a too crude timing resolution to fully characterize the discovered IR pulsations and explain the origin of the IR-optical magnetar emission. Given that pulsations were detected and a lower limit on the pulse fraction was measured, a more optimal observational setup can be used in the future to better characterize the IR pulsations in this source.

\software{Astropy \citep{2013A&A...558A..33A, 2018AJ....156..123A, 2022ApJ...935..167A}, PINT \citep{2021ApJ...911...45L,2024ApJ...971..150S}, CRIMP \url{https://github.com/georgeyounes/CRIMP}, ChatGPT 5.5 Pro Extended \url{https://openai.com/index/introducing-gpt-5-5/}}

\medskip\noindent{\bf Acknowledgments:}
J.H. thanks Zorawar Wadiasingh and Sasha Philippov for useful discussions regarding this work and Han-Long Peng for useful discussions related to their timing model.  We thank the referee for their useful comments which helped improve the clarity of the manuscript.
This work is based on observations made
with the NASA/ESA/CSA JWST. The data were obtained from the Mikulski
Archive for Space Telescopes at the Space Telescope Science
Institute, which is operated by the Association of Universities
for Research in Astronomy, Inc., under NASA contract NAS
5-03127 for JWST. These observations are associated with
program \#2635 and can be accessed via  \dataset[https://doi.org/10.17909/catg-0936]{https://doi.org/10.17909/catg-0936}. Support for program \#2635 was provided by NASA
through a grant from the Space Telescope Science Institute,
which is operated by the Association of Universities for
Research in Astronomy, Inc., under NASA contract NAS
5-03127. 
J.H.\ acknowledges
support from NASA under award number 80GSFC24M0006 and partial support through NuSTAR award NNH21ZDA001N-NUSTAR.
This work made use of Astropy:\footnote{http://www.astropy.org} a community-developed core Python package and an ecosystem of tools and resources for astronomy \citep{2013A&A...558A..33A, 2018AJ....156..123A, 2022ApJ...935..167A}. STScI is operated by the Association of Universities for Research in Astronomy, Inc., under NASA contract NAS5–26555. Support to MAST for these data is provided by the NASA Office of Space Science via grant NAG5–7584 and by other grants and contracts.

\appendix
\label{sec:Appendix}

\section{Time-averaged photometry}
\label{appendix_photometry}
\begin{figure*}
\includegraphics[width=0.42\linewidth]{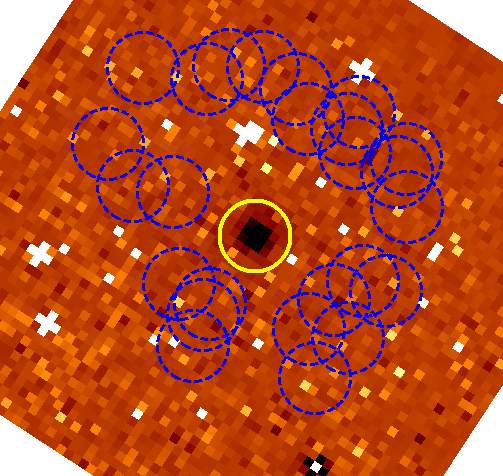}
\hspace{0.5cm}
\includegraphics[width=0.55\linewidth]{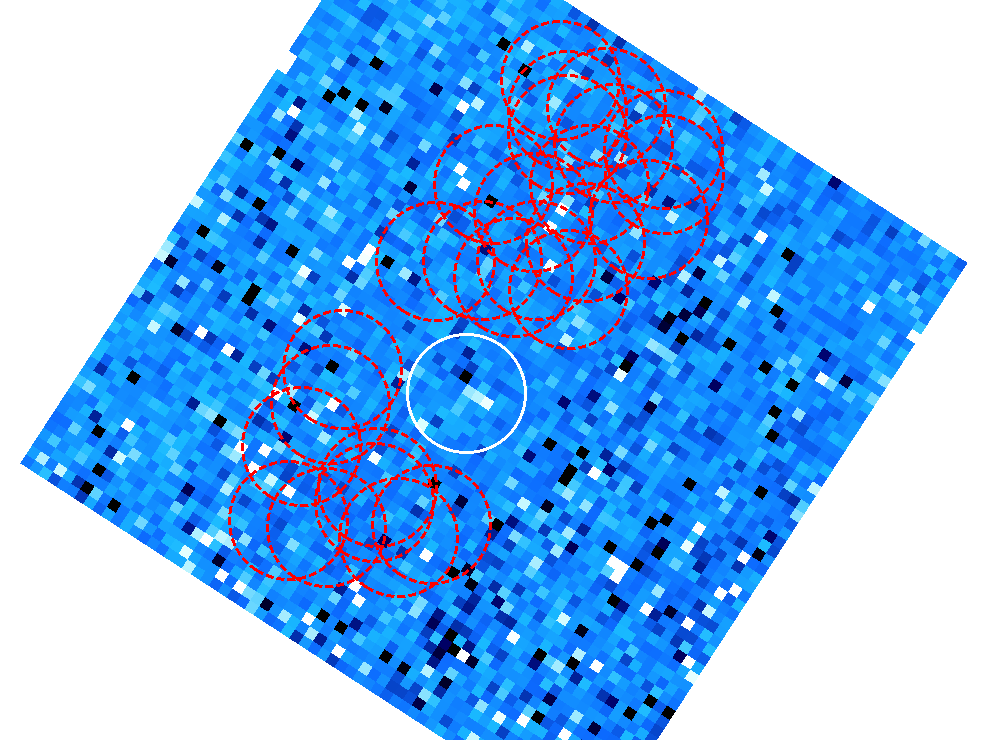}
\caption{Locations of photometry measurement apertures for F410M (left panel) and F070W (right panel). The shown example (maximum size) apertures have radii of $0\farcs25$ 
and $0\farcs17$, respectively. North is up, East to the left.
}
\label{fig:apertures}
\end{figure*}

In Table~\ref{table:psr_photometry}, we list the details of the photometry obtained from the images averaged 
over the 800 ramp integrations in the 
 TSO with filters F410M and F070W and show the locations of the source and background apertures in
Figure~\ref{fig:apertures}.
For the $3\sigma$ upper bound of  F070W, we list
measurement results for 
several aperture radii 
up to 0\farcs17, beyond which a bad pixel gets into the aperture.
The upper bound, however, is relatively stable, 
around 1 $\mu$Jy, within this aperture size range.

We also checked the photometry of the acquisition image which was done in a different filter, F335M, on the $32\times 32$ pixels TA subarray, located at a different part of the B5 chip. 
The image has a bad pixel at $\sim0\farcs06$ from
the target center position. 
Although we excluded 
this bad pixel,
we caution that the derived 
flux density $f_{\rm F335M}= 11.3 \pm 0.4\, \mu$Jy
may be underestimated 
and more uncertain due to the systematics introduced by this bad pixel.

\begin{table*}[h!]
\setlength{\tabcolsep}{0.75em}
\centering          
\begin{tabular}{l c c c c c c c c c c c}    
\hline\hline       
Filter & $\lambda_{\rm p}$ & BW & $g$ &$t_{\rm exp}$ & $r_{\rm extr}$ & $\phi$ & $C_{\rm tot}$ & 
$\overline{C}_b\pm \sigma_{C_b}$
& $\overline{C}_{s}\pm \sigma_{C_s}$ & ${\cal P}_\nu$ & $\langle f_{\nu}\rangle$ or UB\\
& $\mu$m & $\mu$m &  & s & arcsec &  \% & cnts/s & cnts/s & cnts/s & $\mu$Jy\,s/cnts & $\mu$Jy \\
\hline     
F410M & 4.083 & 0.436 & 1.82 & 1806 & 0.156 & 65 & 196.4 & $11.0\pm4.5$ & $185.4 \pm 4.5$ & 0.0809 & $22.9\pm 0.6$\\

F335M & 3.362 & 0.348  & 1.82 & 19.15  & 0.05 & 35 & 80.2 & $30.5 \pm 1.3$ & $49.7 \pm 1.7$ & 0.0791 &  $11.3 \pm 0.4$ \\

F070W & 0.705 & 0.128 & 2.05 & 1806 & 0.09 & 72 & 16.4 & $13.3 \pm 1.6$ &  $3.1 \pm 1.6$ & 0.1034 & $<1.1$ \\

 & & & & & 0.10 & 74 & 18.5 & $16.7 \pm 1.7$ &  $1.8 \pm 1.7$ &  & 
 $<1.0$ \\

 & & & &  & 0.12 & 75 & 23.9 & $23.8 \pm 2.1$ &  $0.2 \pm 2.1$ &  & $<0.9$ \\

 & & & & & 0.14 & 77 & 31.9 & $31.9 \pm 2.3$ &  $0.0 \pm 2.3$ &  & 
 $<0.9$ \\

 & & & & & 0.15 & 78 & 36.9 & $36.7 \pm 2.5$ & $0.2 \pm 2.5$ &  & 
 $<1.0$ \\

 & & & &  &0.17 & 80 & 46.4 & $47.4 \pm 3.5$ & $-1.0 \pm 3.5$ &  & 
 $<1.2$ \\

\hline                
\end{tabular}
\caption{Photometry measurements of the new NIRCam data.
Here $\lambda_{\rm p}$ and BW 
are the pivot wavelength and the 
bandwidth of the filter, $g$ is the gain, $t_{\rm exp}$ is the exposure time, $\phi$ is the fraction of source counts in the aperture with radius $r_{\rm extr}$,
$C_{\rm tot}$ is the total count rate in the source aperture,
$\overline{C}_{\rm b}$ and $\sigma_{C_{\rm b}}$ are the mean and standard deviation of background count rate measurements,
$\overline{C}_{s}=C_{\rm tot}-\overline{C}_b$ and $\sigma_{C_s} = [(\sigma_{C_{\rm b}}^2 + \overline{C}_{s} t_{\rm exp}^{-1} g^{-1}]^{1/2}$ are the net source count rate and its standard deviation,
${\cal P}_\nu$ is the count rate-to-flux conversion factor,
$\langle f_\nu \rangle = (\overline{C}_{s}\pm \sigma_{C_s}) {\cal P}_\nu$/$\phi$
is the mean flux density, and UB $=(\overline{C}_s + 3\sigma_{C_s})\cal{P}_\nu/\phi$ 
is the $3\sigma$ upper bound on the source flux. 
\label{table:psr_photometry} }
\end{table*}

\section{Effects of finite time resolution}
\label{sec:finite_timeres}

In a NIRCam TSO of a periodic source, 
one obtains
a series of source flux values accumulated in equal adjacent time intervals (integration ramps) $\Delta t_{\rm ramp}$, which can be considered as the time resolution. 
Unless 
$\Delta t_{\rm ramp}$ is much shorter than  the source period, the finite time resolution distorts the detected pulsations. 
For instance, 
for purely
sinusoidal pulsations with  frequency $\nu$,
the signal accumulated in the $n$-th 
integration ramp ($n=1, 2, \ldots N$) is proportional to
\begin{equation}
    \frac{1}{\Delta t_{\rm ramp}} \int_{t_0 + (n-1) \Delta t_{\rm ramp}}^{t_0 + n \Delta t_{\rm ramp}} \sin 2\pi\nu t\,dt = \sin 2\pi\nu[t_0 +(n-1/2) \Delta t_{\rm ramp}]\,\, {\rm sinc}(\pi\nu \Delta t_{\rm ramp})\,,
\end{equation}
where $t_0$ is the start time of the first ramp,
and ${\rm sinc}\, x = (\sin x)/x$. 
This equation shows that the integrated signal in each ramp
is proportional to the original  value of the sinusoidal signal in the middle of the 
ramp multiplied by the suppression factor $\xi={\rm sinc}(\pi\nu \Delta t_{\rm ramp})$, which does not depend on  
ramp number. 
This means that the finite time resolution does not change the signal frequency, but it reduces the signal amplitude and pulsed fraction by the reduction factor 
$\xi$,
and reduces the Fourier power of the signal by the factor 
$\xi^2$.
At $\nu \Delta t_{\rm ramp} < 1$ these factors monotonically decrease with increasing $\nu \Delta t_{\rm ramp}$, becoming zero at $\nu \Delta t_{\rm ramp} = 1$.

If the signal contains harmonics $k>1$, then 
the amplitude of the $k$-the harmonic is suppressed by the factor $\xi_k ={\rm sinc} (\pi k\nu \Delta t_{\rm ramp})$, while 
the power of the $k$-th harmonic is suppressed by the factor $\xi_k^2$.
This implies that the finite time resolution not only leads to suppression of the signal amplitude and power, but also distorts its shape.


\begin{thebibliography}{}



\bibitem[Abramkin et al.(2022)]{2022ApJ...924..128A} Abramkin, V., Pavlov, G.~G., Shibanov, Y., et al.\ 2022, \apj, 924, 128. doi:10.3847/1538-4357/ac3a6f

\bibitem[Adams et al.(1987)]{1987ApJ...312..788A} Adams, F.~C., Lada, C.~J., \& Shu, F.~H.\ 1987, \apj, 312, 788. doi:10.1086/164924

\bibitem[Archibald et al.(2017)]{2017ApJ...834..163A} Archibald, R.~F., Kaspi, V.~M., Scholz, P., et al.\ 2017, \apj, 834, 2, 163. doi:10.3847/1538-4357/834/2/163

\bibitem[Astropy Collaboration et al.(2022)]{2022ApJ...935..167A} Astropy Collaboration, Price-Whelan, A.~M., Lim, P.~L., et al.\ 2022, \apj, 935, 2, 167. doi:10.3847/1538-4357/ac7c74

\bibitem[Astropy Collaboration et al.(2018)]{2018AJ....156..123A} Astropy Collaboration, Price-Whelan, A.~M., Sip{\H{o}}cz, B.~M., et al.\ 2018, \aj, 156, 3, 123. doi:10.3847/1538-3881/aabc4f

\bibitem[Astropy Collaboration et al.(2013)]{2013A&A...558A..33A} Astropy Collaboration, Robitaille, T.~P., Tollerud, E.~J., et al.\ 2013, \aap, 558, A33. doi:10.1051/0004-6361/201322068


\bibitem[Bushouse et al.(2025)]{2025zndo..17515973B} Bushouse, H., Eisenhamer, J., Dencheva, N., et al.\ 2025, Zenodo, 1.20.2. doi:10.5281/zenodo.17515973


\bibitem[Chu et al.(2026)]{2026ApJS..284...25C} Chu, C.-Y., Hu, C.-P., Enoto, T., et al.\ 2026, \apjs, 284, 1, 25. doi:10.3847/1538-4365/ae5655 


\bibitem[Dhillon et al.(2005)]{2005MNRAS.363..609D} Dhillon, V.~S., Marsh, T.~R., Hulleman, F., et al.\ 2005, \mnras, 363, 609. doi:10.1111/j.1365-2966.2005.09465.x


\bibitem[Dhillon et al.(2009)]{2009MNRAS.394L.112D} Dhillon, V.~S., Marsh, T.~R., Littlefair, S.~P., et al.\ 2009, \mnras, 394, 1, L112. doi:10.1111/j.1745-3933.2009.00623.x

\bibitem[Dhillon et al.(2011)]{2011MNRAS.416L..16D} Dhillon, V.~S., Marsh, T.~R., Littlefair, S.~P., et al.\ 2011, \mnras, The first observation of optical pulsations from a soft gamma repeater: SGR 0501+4516, 416, 1, L16. doi:10.1111/j.1745-3933.2011.01088.x


\bibitem[Dib \& Kaspi(2014)]{2014ApJ...784...37D} Dib, R. \& Kaspi, V.~M.\ 2014, \apj, 784, 37. doi:10.1088/0004-637X/784/1/37



\bibitem[Durant \& van Kerkwijk(2006)]{2006ApJ...652..576D} Durant, M. \& van Kerkwijk, M.~H.\ 2006, \apj, 652, 576. doi:10.1086/507605


\bibitem[Ertan \& Cheng(2004)]{2004ApJ...605..840E} Ertan, \"{U}., \& Cheng, K.~S.\ 2004, \apj, 605, 840. doi:10.1086/382502

\bibitem[Ertan et al.(2007)]{2007ApJ...657..441E} Ertan, {\"U}., Erkut, M.~H., Ek{\textcommabelow s}i, K.~Y., et al.\ 2007, \apj, 657, 1, 441. doi:10.1086/510303


\bibitem[Gavriil et al.(2008)]{2008AIPC..983..234G} Gavriil, F.~P., Dib, R., \& Kaspi, V.~M.\ 2008, 40 Years of Pulsars: Millisecond Pulsars, Magnetars and More, 983, 234. doi:10.1063/1.2900150


\bibitem[Gendreau et al.(2016)]{2016SPIE.9905E..1HG} Gendreau, K.~C., Arzoumanian, Z., Adkins, P.~W., et al.\ 2016, \procspie, The Neutron star Interior Composition Explorer (NICER): design and development, 9905, 99051H. doi:10.1117/12.2231304


\bibitem[Gordon et al.(2022)]{2022AJ....163..267G} Gordon, K.~D., Bohlin, R., Sloan, G.~C., et al.\ 2022, \aj, 163, 6, 267. doi:10.3847/1538-3881/ac66dc


\bibitem[Guillot et al.(2019)]{Guillot2019} Guillot, S., Pavlov, G.~G., Reyes, C., et al.\ 2019, \apj, 874, 2, 175. doi:10.3847/1538-4357/ab0f38

\bibitem[Hare et al.(2021)]{2021ApJ...923..249H} Hare, J., Volkov, I., Pavlov, G.~G., et al.\ 2021, \apj, 923, 2, 249. doi:10.3847/1538-4357/ac30e2

\bibitem[Hare et al.(2024)]{2024ApJ...972..176H} Hare, J., Pavlov, G.~G., Posselt, B., et al.\ 2024, \apj, 972, 176. doi:10.3847/1538-4357/ad5ce5

\bibitem[Harrison et al.(2013)]{2013ApJ...770..103H} Harrison, F.~A., Craig, W.~W., Christensen, F.~E., et al.\ 2013, \apj, 770, 103. doi:10.1088/0004-637X/770/2/103


\bibitem[Hulleman et al.(2004)]{2004A&A...416.1037H} Hulleman, F., van Kerkwijk, M.~H., \& Kulkarni, S.~R.\ 2004, \aap, 416, 1037. doi:10.1051/0004-6361:20031756


\bibitem[Kashyap et al.(2010)]{Kashyap2010} Kashyap, V.~L., van Dyk, D.~A., Connors, A., et al.\ 2010, \apj, 719, 1, 900. doi:10.1088/0004-637X/719/1/900


\bibitem[Kaspi \& Beloborodov(2017)]{2017ARA&A..55..261K} Kaspi, V.~M. \& Beloborodov, A.~M.\ 2017, \araa, 55, 1, 261. doi:10.1146/annurev-astro-081915-023329


\bibitem[Kern \& Martin(2002)]{2002Natur.417..527K} Kern, B. \& Martin, C.\ 2002, \nat, 417, 527. doi:10.1038/417527a


\bibitem[Luo et al.(2021)]{2021ApJ...911...45L} Luo, J., Ransom, S., Demorest, P., et al.\ 2021, \apj, 911, 1, 45. doi:10.3847/1538-4357/abe62f


\bibitem[Mu{\~n}oz-Darias et al.(2016)]{2016MNRAS.458L.114M} Mu{\~n}oz-Darias, T., de Ugarte Postigo, A., \& Casares, J.\ 2016, \mnras, 458, L114. doi:10.1093/mnrasl/slw024

\bibitem[Olausen \& Kaspi(2014)]{2014ApJS..212....6O} Olausen, S.~A. \& Kaspi, V.~M.\ 2014, \apjs, 212, 6. doi:10.1088/0067-0049/212/1/6

\bibitem[Pavlov et al.(2021)]{2021jwst.prop.2635P} Pavlov, G.~G., Kargaltsev, O., Hare, J., \& Posselt, B. 2021, JWST Proposal. Cycle 1, ID. \#2635

\bibitem[Peng et al.(2026)]{2026ApJ...999...85P} Peng, H.-L., Weng, S.-S., Ge, M.-Y., et al.\ 2026, \apj, 999, 1, 85. doi:10.3847/1538-4357/ae4007


\bibitem[Rieke et al.(2023)]{2023PASP..135b8001R} Rieke, M.~J., Kelly, D.~M., Misselt, K., et al.\ 2023, \pasp, 135, 028001. doi:10.1088/1538-3873/acac53


\bibitem[Shaw et al.(2025)]{2025AJ....169...21S} Shaw, A.~W., Kaplan, D.~L., Gandhi, P., et al.\ 2025, \aj, 169, 1, 21. doi:10.3847/1538-3881/ad8eb1

\bibitem[Susobhanan et al.(2024)]{2024ApJ...971..150S} Susobhanan, A., Kaplan, D.~L., Archibald, A.~M., et al.\ 2024, \apj, 971, 2, 150. doi:10.3847/1538-4357/ad59f7


\bibitem[Tendulkar et al.(2015)]{2015ApJ...808...32T} Tendulkar, S.~P., Hasc{\"o}et, R., Yang, C., et al.\ 2015, \apj, 808, 32. doi:10.1088/0004-637X/808/1/32

\bibitem[VanderPlas(2018)]{2018ApJS..236...16V} VanderPlas, J.~T.\ 2018, \apjs, 236, 1, 16. doi:10.3847/1538-4365/aab766


\bibitem[Wang et al.(2006)]{2006Natur.440..772W} Wang, Z., Chakrabarty, D., \& Kaplan, D.~L.\ 2006, \nat, 440, 772. doi:10.1038/nature04669

\bibitem[Zane et al.(2011)]{2011ASSP...21..329Z} Zane, S., Nobili, L., \& Turolla, R.\ 2011, High-Energy Emission from Pulsars and their Systems, 21, 329. doi:10.1007/978-3-642-17251-9\_26

\end{thebibliography}
\end{document}